\documentclass[12pt]{article}
\pdfoutput=1
\usepackage{amssymb}
\usepackage{amsmath}
\usepackage{amsfonts}
\usepackage{graphicx}
\usepackage{mathrsfs}
\usepackage{comment}
\usepackage{cite}
\usepackage{hyperref}
\usepackage{color}

\newcommand{\Slash}[1]{{\ooalign{\hfil#1\hfil\crcr\raise.167ex\hbox{/}}}}

\newcommand{\ket}[1]{ | {#1} \rangle }
\newcommand{\bef}{\begin{figure}}  \newcommand{\eef}{\end{figure}}
\newcommand{\bec}{\begin{center}}  \newcommand{\eec}{\end{center}}
  
\newcommand{\laq}[1]{\label{eq:#1}}  

\newcommand{\Eq}[1]{Eq.(\ref{eq:#1})}

\newcommand{\eq}[1]{(\ref{eq:#1})}

\newcommand{\ab}[1]{\left|{#1}\right|}

\newcommand{\SU}[1]{{\rm SU{#1} } }

\def\({\left(}
\def\){\right)}

\def\O{\mathcal{O}}
\def\U{\mathop{\rm U}}

\def\tr{\mathop{\rm tr}}
\def\ebq{\end{equation} \begin{equation}}
\newcommand{\OR}{~{\rm or}~}
\newcommand{\AND}{~{\rm and}~}

\newcommand{\GEV}{ {\rm \, GeV} }
\newcommand{\TEV}{ {\rm \, TeV} }
\def\o{\over}
\def\a{\alpha}
\def\b{\beta}

\def\d{\delta}
\def\e{\epsilon}
\def\f{\phi}

\def\k{\kappa}
\def\l{\lambda}

\def\r{\rho}

\def\D{\Delta}
\def\G{\Gamma}

\def\L{\Lambda}
\def\F{\Phi}

\def\tl{\tilde}
\def\*{\dagger}

\begin{document}

\begin{center}

\hfill KEK--TH--2179

\vspace{1.5cm}

{\Large\bf Direct baryogenesis in the broken phase}
\vspace{1.5cm}

{\bf Takehiko Asaka$^{(a)}$, Hiroyuki Ishida$^{(b)}$, Wen Yin$^{(c,d)}$}

{$^{\rm (a)}${\it 
Department of Physics, Niigata University, Niigata 950-2181, Japan}

\vskip 0.1in

$^{\rm (b)}${\it
Theory Center, IPNS, KEK, Tsukuba, Ibaraki 305-0801, Japan}

\vskip 0.1in

$^{\rm (c)}${\it Department of Physics, KAIST, Daejeon 34141, Korea}
}

\vskip 0.1in

$^{\rm (d)}${\it Department of Physics, Faculty of Science, The University of Tokyo, 
} \\
\small{\em Bunkyo-ku, Tokyo 113-0033, Japan}\\[0.5ex]  

\vspace{12pt}
\vspace{1.5cm}

\date{\today $\vphantom{\bigg|_{\bigg|}^|}$}

\abstract{
We show a new mechanism for baryogenesis 
where the reheating temperature can be smaller than the electroweak scale. 
The baryon number symmetry is violated by a dimension nine operator which conserves a baryon parity. 
A high energy quark from the decay of a heavy particle, 
{\it e.g.} inflaton, modulus or gravitino, undergoes flavor oscillation, 
and is thermalized due to the scatterings with the ambient thermal plasma. 
We point out that the baryon asymmetry of our universe can be generated due to the scatterings via the baryon number violating operator.  
Our scenario can be tested in neutron-antineutron oscillation experiments
as well as other terrestrial experiments.
}
\end{center}
\clearpage

\setcounter{page}{1}

\setcounter{footnote}{0}
\section{Introduction}

The baryon asymmetry of the universe is one of the leading mysteries of the inflationary cosmology. 
Depending on the reheating temperature, $T_R$, 
various baryogenesis scenarios are studied in different contexts.
For $10^{12}\GEV \gtrsim T_R\gtrsim 100\GEV$, where the baryon number violation by a sphaleron process is effective, 
the baryon asymmetry can be transferred from the lepton asymmetry generated at the high temperatures.
For instance, in thermal leptogenesis~\cite{Fukugita:1986hr}, 
 the lepton asymmetry is produced by the CP-violating decay of a heavy right-handed neutrino, 
which requires $T_R\gtrsim 10^{8-9}\GEV.$
For $T_R\lesssim 100\GEV$, on the other hand, 
the construction of a successful baryogenesis scenario becomes difficult.
This is because we need a baryon number violating process while the stability of proton must be guaranteed. 
Such an attempt to realize low scale baryogenesis was initiated 
by Ref.~\cite{Dimopoulos:1987rk} in a context of $R$-parity violation in supersymmetric theories. 
(Also, see,{\it e.g.}, Refs.~\cite{Barbier:2004ez, Babu:2006xc, Grojean:2018fus, Pierce:2019ozl} and references therein.)%
\footnote{Asymmetric dark matter~\cite{Barr:1990ca, Kaplan:1991ah, Kitano:2004sv, Kaplan:2009ag, Allahverdi:2017edd}, or
baryogenesis before the last period of reheating~\cite{Davidson:2000dw} can also explain the baryon asymmetry in this temperature range.}

It becomes clear that a field once dominates over the universe, 
and then decays into the standard model (SM) particles to reheat the universe. 
Such a scalar should exist to drive inflation: the inflaton. 
In addition to the inflaton, moduli, axions, or gravitino may play the role. 
In fact, many underlying theories, such as string theory and M-theory, 
predict the existence of scalars and fermions coupled to the SM particles as weak as gravity.
If those particles are produced in the early universe, 
they dominate over the universe. The decay then leads to a low reheating temperature 
due to the weakness of the coupling constants. 
The temperature can be easily lower than the electroweak (EW) scale, 
$T_R\lesssim 100\GEV$, and hence the baryogenesis becomes difficult.

In this paper, we propose a baryogenesis scenario at $T_R \lesssim 100\GEV$. 
We show that the baryon asymmetry can be directly generated 
from the perturbative decay of a heavy particle, which could be the inflaton etc, weakly coupled to the SM particles. 
We introduce a baryon number violating but baryon parity preserving operator. 
In an effective theory for the SM particle contents, a dimension nine baryon number violating operator plays the role.
Thanks to the baryon parity, the proton is stable. 
Energetic quarks produced by the heavy particle undergo flavor oscillation 
due to the misalignment of the bases of the quark masses and the interactions. 
The oscillation can cause CP violation analogous to the ordinary neutrino oscillation.
The baryon asymmetry is created by the first scattering via the dimension nine operator. 
This scenario can be tested, {\it e.g.}, in neutron-antineutron oscillation experiments~%
\cite{Phillips:2014fgb,Milstead:2015toa,Frost:2016qzt,Hewes:2017xtr}. 
By building a simple renormalizable UV model, we show our scenario works as well.  
In the case, our scenario may also have implications on flavor changing neutral currents (FCNCs) and CP-violating processes.

In the context of quantum oscillation,  
the possible baryogenesis scenarios at $T_R\lesssim 100\GEV$ were studied in terms of hadrons at the confinement phase~\cite{McKeen:2015cuz, Aitken:2017wie, Elor:2018twp}. 
In particular, the baryogenesis by the heavy baryon-antibaryon oscillation 
has been shown to be possible~\cite{McKeen:2015cuz, Aitken:2017wie}. 
It was found that the baryon asymmetry production 
can be efficient, and is consistent with the experimental constraints by enhancing relevant dimension nine operators with (approximate) flavor symmetry. 

In contrast, we focus on the energetic quarks produced from the decays of very heavy particles before the confinement.
The quark flavor oscillation happens {\it a la} the baryogenesis via the right-handed neutrino oscillation~\cite{Akhmedov:1998qx,Asaka:2005pn,Asaka:2011wq}.%
\footnote{Here right-handed neutrinos are produced from the scatterings of the particles in the thermal bath.
The null total lepton asymmetry is separated into the right- and left-handed neutrino sectors 
as the same amount but the opposite sign 
where the latter one is transferred into the baryon asymmetry by sphaleron. 
Interestingly
the reheating temperature is allowed to be as low as $\O(100)\GEV$.
} 
In particular, we point out that the baryon asymmetry 
can be significantly produced due to the quark-plasma scatterings at a high center-of-mass energy. 
The mechanism allows us to have weak enough relevant operators without conflicting with experimental constraints.

A similar mechanism has been considered in the context of active neutrino oscillation 
with higher dimensional terms for $T_R\gtrsim 10^8\GEV$~\cite{Hamada:2016oft}.
It was shown that baryogenesis is possible in the SM 
plus the dimension five Majorana neutrino mass term, $LLHH$, 
explaining the ordinary neutrino oscillation~\cite{Hamada:2018epb}. 
The key point is that the leptons, produced from the inflaton decays or the scatterings of the plasma, 
undergo flavor oscillation due to the thermal masses. (See also Ref.~\cite{Eijima:2019hey} for the case with a light right-handed neutrino.)

\section{Baryon asymmetry from quark flavor oscillation}


To discuss our mechanism, we first consider an effective theory made up by the SM particle contents 
where all the quarks are charged under the parity. The leading operators changing the baryon number are
\begin{equation}
\laq{1}
{\cal L}\supset \k_1 Q^4(d^*)^2+\k_2 u^2 d^4+\k_3 (Q^*)^2 d^3 u+h.c.\,,
\end{equation}
where $Q:~ (1/6, 2,3)$ and $u: ~(-2/3,1,\bar{3})$, $d: ~(1/3,1,\bar{3})$ denote left-handed quarks and right-handed anti-quarks, respectively, in Weyl notation, with the corresponding representations under the SM gauge group: $(\U(1)_Y,\SU(2)_L,\SU(3)_c)$.
Here $\k_{1,2,3} $ are the couplings with dimension $-5$, 
we have omitted the flavor, Lorentz, and gauge indices.  
Although the baryon number symmetry is violated by two units, proton remains stable due to the $Z_2$ symmetry where 
all the quarks are odd but the other fields are even. We omit the contribution of dimension six operators which conserve the baryon number. The dimension six operators are not important unless they change drastically the thermalization process (see Sec.~\ref{sec:4}).

The highest energy scale, $\L_{\rm cutoff}$, of the model justifying the perturbative expansion is obtained as
\begin{equation}
\laq{cutoff}
\(\frac{1}{ 16\pi^2}\)^4\O(|\k_{1,2,3}|^2)\L_{\rm cutoff}
^{10}\lesssim 1\,.
\end{equation}
The center-of-mass energy of a scattering process for a quark should satisfy 
\begin{equation}
\laq{emax}
E_{\rm cm} \lesssim \Lambda_{\rm cutoff} \simeq (4\pi)^{4/5} \O(|\k_{1,2,3}|^{-1/5})\,.
\end{equation}
Before the energy scale $E_{\rm cm}$ becomes around $\L_{\rm cutoff},$
the effective theory may be replaced by a UV renormalizable model.\footnote{Alternatively it can just become non-perturbative.} A UV renormalizable model will be discussed in Sec.~\ref{sec:4}.
In what follows, we use the effective theory to describe our scenario
where our discussion should be applied to all the UV models with heavy enough new states. 

\subsection{Mechanism}
Suppose that a weakly coupled heavy particle, $\f$, with mass $m_{\rm \f}$ 
decays into the SM particles including quark/antiquark. 
Here $\f$ is a scalar particle, {\it e.g.} an inflaton, a modulus or an axion. 
Also our discussion can be extended to the fermion case, {\it e.g.} $\f$ is a gravitino, straightforwardly. 
For a while before the decay, 
the energy density of $\phi$ is assumed to dominate the universe 
due to the longevity caused by the weak coupling. 
Then the out of equilibrium decays with the total width $\G_{\rm \f}$ reheat the universe. 
The reheating temperature is given as 
$T_R \equiv  \(g_* \pi^2 /90\)^{-1/4} \sqrt{\G_{\f} M_{\rm p}}$ 
with $M_{\rm p}\approx 2.4\times 10^{18}\GEV$ being the reduced-Planck scale 
and $g_*$ being the relativistic degrees of freedom. 
Here we focus the region of the reheating temperature
\begin{equation}
\laq{range}
  ~1\GEV\lesssim T_R \lesssim 100\GEV%
\end{equation}
for simplicity. 
At the range of temperature the effects of confinement can be omitted. 
We will mention the application of our mechanism out of this range later.

At the moment of an inflaton decay at $t= t_R\simeq 1/\G_{\rm \f}$,
two components constitute the universe, 
the energetic quarks with energy $\sim  m_\f/2\gg T_R,$%
\footnote{We have assumed a two-body decay for simplicity, but it is not necessary. 
As long as the energy of emitted quark is of the order $m_\f$ our prediction does not change much.}
and the thermal plasma characterized by the temperature of $T_R$. 
The latter component is produced due to the decays of $\f$ at $t<t_R$ 
and is diluted by the entropy production of the decaying $\f$. The products are soon thermalized and form the thermal bath. 
The former component is from the direct decay at this moment, which is injected into the thermal bath. 
The energetic quarks soon scatter with ambient thermal plasma, 
dissipate the energy, and are thermalized in the end. This thermalization process is obviously a one-way process, during which the baryon asymmetry can be created. In the following, we concentrate on this process. 

Let us track a quark quantum state during the thermalization process. 
For instance, a state of an up-type quark is written as 
$\ket{ U_{\rm \f}}_{t=t_R}$ at $t=t_R$. 
This quantum state of the quark can be expanded by the mass eigenstates as follows 
\begin{equation}
\ket{ U_{\rm \f}}|_{t=t_R}= V_u^P \ket{u} + V_c^P \ket{c} + V_t^P \ket{t}\,,
\end{equation}
where 
\begin{equation}
V_i^P \equiv \langle i\ket{U_{\rm \f}}\,,
\end{equation}
with $i$ being $u,c,t$.
Before the first scattering the state undergoes flavor oscillation 
and
at $t=t_R+\D t$ the state reads 
\begin{equation}
\laq{osc}
\ket{ U_{\rm\f}}|_{t=t_R+\Delta t} 
= 
V_u^P \exp{\(i \frac{m_u^2}{m_{\f}}\Delta t \)}\ket{u} 
+ 
V_c^P \exp{\(i \frac{m_c^2}{m_{\f}}\Delta t \)} \ket{c} 
+ 
V_t^P \exp{\(i \frac{m_t^2}{m_{\f}}\Delta t \)}\ket{t}\,,
\end{equation}
where we 
define the states $\ket{u,c,t}$ with including the flavor blind phase of $\exp[-i m_{\f}t/2+...]$ 
which does not contribute to the flavor oscillation.

The energetic quark can not travel freely (or coherently) for a long time 
because of the preexisting thermal plasma. 
The flavor oscillation is terminated due to the scattering with the ambient plasma at the time scale 
\begin{equation}
\laq{th}
(\Delta t)^{-1}\equiv \Gamma_{\rm th}\,.
\end{equation} 
Here $\Gamma_{\rm th}$ is the thermalization rate obtained 
from the inelastic scatterings between the quark and the plasma.%
\footnote{The decay rate of the top quark with a boost factor is $\propto \alpha_2 m_t^2/m_\f$, 
which will be slower than the thermalization processes shown below.} 
The quark scatters with plasma to produce many soft gluons via a Bremsstrahlung emission, whose rate is $\propto T.$
In the dense medium, the multi-gluon emission rate is suppressed by a so-called Landau-Pomeranchuk-Migdal (LPM) effect~\cite{Landau:1953um,Migdal:1956tc}, due to which 
the amplitudes of scattering cancels among different diagrams unless the momentum transfer becomes large enough. 
This effect suppresses the energy loss rate by $\sqrt{2T/m_\f}$ compared with the one derived 
from a na\"{i}ve inelastic scattering (See Refs.~\cite{Allahverdi:2002pu, Kurkela:2011ti, Harigaya:2013vwa, Harigaya:2014waa} for "bottom-up" thermalization): 
\begin{equation}
\Gamma_{\rm LPM} \simeq C' \alpha_3^2 T_R \sqrt{\frac{2T_R}{m_\f}}\,,
\end{equation}
where $C'=\O(1)$ denotes the theoretical uncertainty of the thermalization rate.%

An important observation in our scenario is that other than the gluon propagating interaction, 
the dimension nine operator also contributes to the scattering, as a $2\to 4$ process. 
The $2 \to 4$ scattering rate is given by 
\begin{equation}
\laq{BV}
\G_{\rm BV}= \frac{C(\k_1,\k_2,\k_3)}{4\pi\cdot (16\pi^2)^2}\frac{E_{\rm cm}^8}{\L^{10}}\times \frac{3\zeta{(3)}T_R^3}{2\pi^2}\,.
\end{equation}
Here, we have explicitly extracted the typical scale of $\kappa_i$ by denoting $\Lambda$,  $4 \pi \cdot (16\pi^2)^2$ is the phase space suppression ({\it cf.} Ref. \cite{Asaka:1994bx}), and the last factor denotes the number density of a thermalized fermion (of two spin components).
The center-of-mass energy is approximated as 
\begin{equation} 
E_{\rm cm}\sim \sqrt{  T_R m_\f }\,. 
\end{equation}
$C$ is a dimensionless function of $\k_{1,2,3}$ 
which can be calculated from the imaginary part of the self-energy 
 of the up-type quarks in the thermal environment~\cite{Bellac:2011kqa}. 
For instance,  we can consider the interaction of the form $(\k_2)^{ j_1j_2k_1 k_2k_3k_4} \epsilon_{\a_1\a_2\a_3}\epsilon_{\b_1 \b_2 \b_3 } u^{\a_1}_{j_1}d^{\a_2}_{k_1}d^{\a_3}_{k_2}u^{\b_1}_{j_2}d^{\b_2}_{k_3}d^{\beta_3}_{k_4}$, where we have explicitly shown the indices of flavor $j_a,k_a=1,2,3$, denoting the generation in the mass basis, and color $\a_a,\b_a=1,2,3.$  We assume for simplicity that $j_a$ as well as $k_a$ are symmetric for different $a$ in $\k_2$. In the case one obtains \begin{equation} C\equiv v_l^* v_m C_{lm}\end{equation}
 where 
\begin{equation}
C_{l m}\sim 2^8 \L^{10} \sum_{j_2}\sum_{ k_1\geq k_2\geq k_3\geq k_4}{(\k_{2})_{ l j_2 k_1 k_2 k_3 k_4}^* (\k_{2})_ {m j_2 k_1 k_2 k_3 k_4}}.
\end{equation}
$C_{l m}$ is related with a two point function of flavor $l$ and $m$ quarks. 
$v_l$ is the norm $1$ eigenvector of $C_{lm}$. 
$v_l$ will define the interaction basis, which is important for the flavor to be ``observed."
Here, $2^8$ comes from the color factors, permutations of the flavor indices and contractions of spin indices.
If one assume that $\k_2$ with arbitrary indices are of order $\L^{-5},$ the summation provides $\sim 3\times 6!/(2! 4!)\L^{-10}\sim 45 \L^{-10}.$
One obtains 
\begin{equation}
C= \O(10^4)\,.
\end{equation}
If there are other contributions from $\k_1,\k_3$, the value can be even larger. 
On the other hand, if only a single set of indices of $(\k_{2})$ dominates, which happens if there is a specific flavor structure, one obtains $C\sim 2^8.$

One can see that this scattering is efficient for the energetic quark 
but it becomes inefficient after the quark loses its energy. 
If the center-of-mass energy is high enough, the scattering also contributes dominantly to the thermalization over the energy loss process. 
It turns out that the thermalization rate can be estimated as 
\begin{equation}
\laq{th}
\G_{\rm th}\simeq \max{\(\G_{\rm LPM}, \G_{\rm BV}\)}\,.
\end{equation}

In the case the $2\to 4$ process is flavor dependent, 
 the flavor can be ``observed." 
Suppose that the ``observed" state by a $2 \to 4$ process is $\ket R$ which is an eigenstate of the interaction basis satisfying $v^*_i= \langle i\ket R.$
Then the difference between the probability for producing a quark state $\ket R$ and its CP conjugate propability is given as 
\begin{equation}
\laq{CPV}
P_{U_{\f}\to R}-P_{\bar{U}_{\f}\to \bar{R}} 
\simeq 
4 \sum_{j\geq k}\Im[ V^P_j v_j^* v_k (V^P_k)^*] \sin{\( {m_{k}^2- m_j^2\o m_{\f}} \D t \)}\,.
\end{equation}
One finds if either $V^P_j$ or $v_j$ contains a CP-odd phase, the CP violation probability can be non-zero. 
The exception is that either $V^P_i=\delta_{i a}$ or $v_j=\delta_{j a}$ 
is aligned to the quark mass basis, 
where $a$ denotes an index of the mass eigenstate.

Since the interaction violates the baryon number by two units, 
all the Sakharov's conditions~\cite{Sakharov:1967dj} can be satisfied 
once the flavor oscillation is terminated by the $2 \to 4$ process. 
Consequently, the baryon asymmetry can be generated. 
The ``observation" happens for the fraction 
$\simeq \G_{\rm BV}/\G_{\rm th}$
of the total energetic quarks and antiquarks produced by the $\f$ decays.

Now we can estimate the produced amount of the baryon asymmetry at the first scattering. 
The generated baryon to entropy ratio can be given by 
\begin{align}
\laq{as}
{\D_B\o s} 
&\simeq 
{3T_R\o 4m_\f} B \times (P_{U_{\f}\to R}-P_{\bar{U}_{\f}\to \bar{R}}) 
\times 2{\G_{ \rm BV}\over \G_{\rm th}}\notag\\
&\simeq 
9\times 10^{-10} B \xi_{CP}C'^{-2}   \(\frac{C}{10^4}\) 
 \({E_{\rm cm} \over 2\L  }\)^6 \({T_R\o 90\GEV}\)^2 \({200\TEV\o \L}\)^4\,,
\end{align}
where $s$ is the entropy density of the universe and we define 
\begin{equation}
\xi_{CP}\equiv \sum_{k=c,u }\Im[V^P_t v^*_t v_k (V^P_k)^* ]\,.
\end{equation}
$B$ is the decay branching ratio to the quark states characterized by $\ket{U_{\f}}$. 
We have used $m_t^2/m_\f \times \G^{-1}_{\rm th}\sim m_t^2 \a_s^{-2}  (E_{\rm cm}^2 T_R^2)^{-1/2}\ll 1$ to expand the $\sin$ function in \Eq{CPV}.
The factor of 2 in front of $\G_{\rm BV}/\G_{\rm th}$ is from the ``two" unit violation of baryon number 
by the $2\to 4$ process. 
In the second row, we pick up the contribution of the top quark since $m_t$ is the largest quark mass.

\subsection{Predictions}
Notice that although \Eq{as} increases as $E_{\rm cm}$ increases, 
there is an upper bound on the generated baryon asymmetry. 
This is because when $\G_{\rm LPM}<\G_{\rm BV}$, {\it i.e.}
\begin{align}
E_{\rm cm} \gtrsim \bar{E} 
\simeq  
250\TEV \times 
\left( \frac{10^4}{C /C'} \right)^{1/9} \(\frac{\L}{100\TEV}\)^{10/9} \(\frac{100\GEV}{T_R}\)^{1/9}\,,
\end{align}
the dominant energy dissipation occurs via $2 \to 4 $ scattering process and $\Delta t\sim \G^{-1}_{\rm BV}$ decreases as $E_{\rm cm}$ increases. 
Thus the asymmetry generated by the oscillation tends to decrease. 
However when $\bar{E}\gg \L_{\rm cutoff}$ our effective treatment of dimension nine operator becomes invalid, and we consider that
there is a UV renormalizable theory above the scale, where the $\G_{\rm BV}$ is suppressed by the center-of-mass energy from dimensional grounds. (See Sec.\,\ref{sec:4}.)\footnote{This is also the 
reason we do not consider the scattering between two energetic quarks which have the center-of-mass energy $m_\f\gg \L$.  }

\begin{figure}[t]
\begin{center}  
\includegraphics[width=74mm]{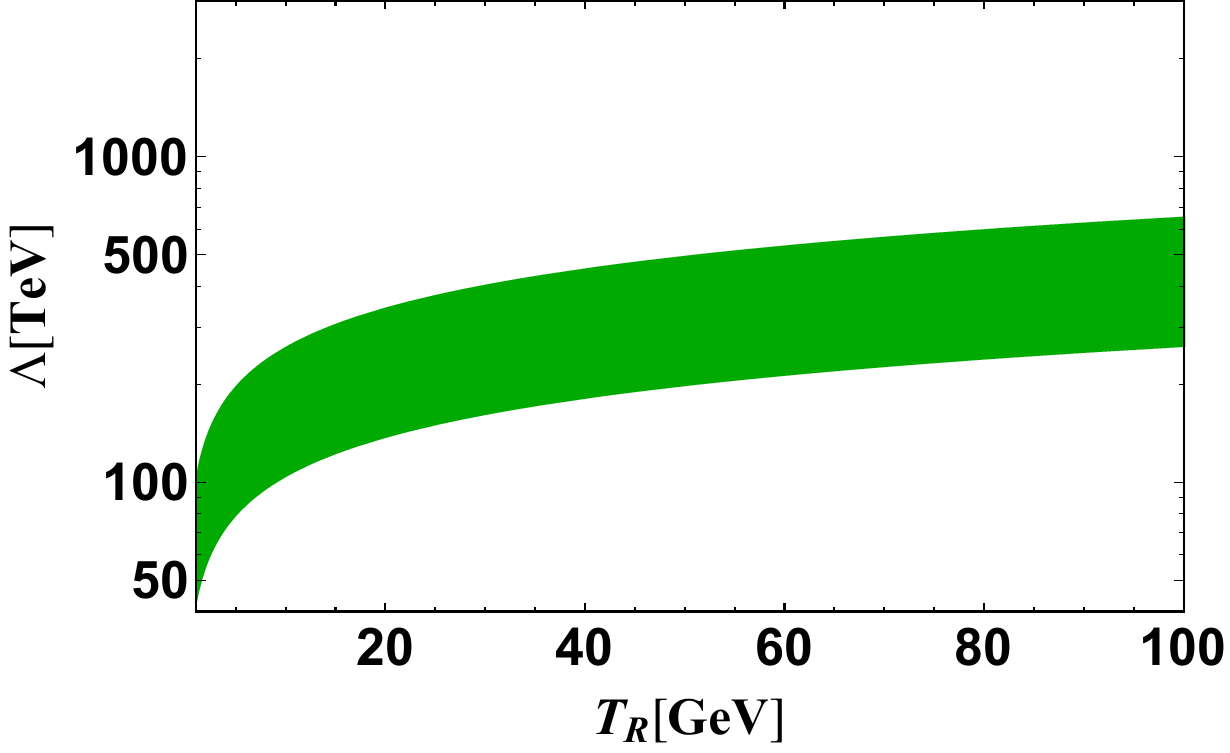}
\includegraphics[width=74mm]{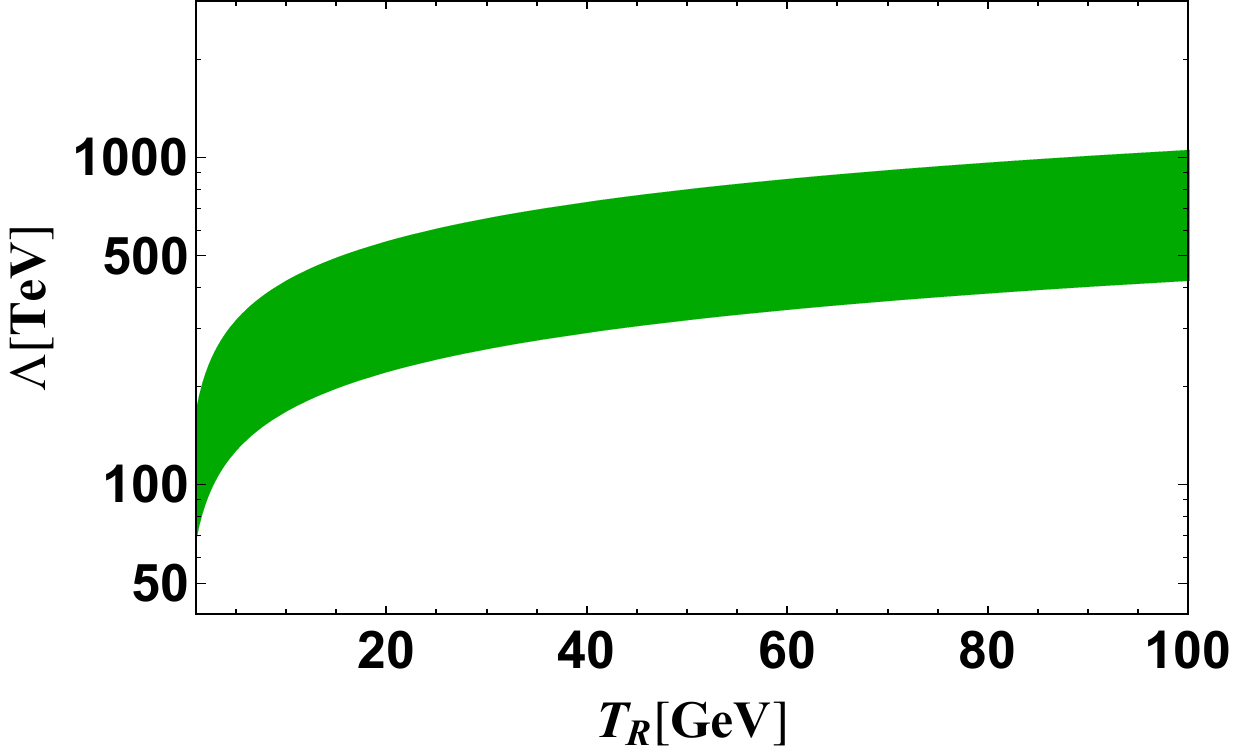}
\end{center}
\caption{The analytical upper bound on $\L~[\TEV]$, the scale of the dominant dimension nine interaction, 
to get correct amount of the baryon asymmetry as a function of the reheating temperature. 
We take $C=2^8$ (left) and $C=5\times 10^4$ (right). In both figures, $B\xi_{CP}=1$ is taken for simplicity, and 
 the range corresponds to the variation of $C'=[0.3, 3]$. }
\label{fig:1}
\end{figure}

We can estimate the maximal amount of the asymmetry generated due to the first scattering 
by setting $E_{\rm cm}=\bar{E}$ in \Eq{as} as 
\begin{equation}
\laq{maxas}
{\D_B^{\rm max} \o s} 
\simeq 
7.8\times 10^{-11} B \xi_{CP} C'^{-4/3} \ab{\frac{C}{10^4}}^{1/3}  \({600\TEV \over  \L  }\)^{10/3} \({T_R\o 100\GEV}\)^{4/3}\,. 
\end{equation}
In Fig.~\ref{fig:1}, we show the upper bound of $ \L$ in the $T_R \mathchar`- \L$ plane with $B \xi_{CP}=1$ for simplicity 
to get a correct amount of the baryon asymmetry that is measured as~\cite{Aghanim:2018eyx}
\begin{equation}
{\D^{\rm obs}_B\o s}\simeq 8.7\times10^{-11}\,.
\end{equation}
We take $C=2^8 \AND 5\times 10^4$ in the left and right panels, respectively. 
In both panels, $C'=[0.3, 3]$ is varied to take account of the theoretical uncertainty which is dominantly from the LPM effect. 
The predicted region compatible with the observed value of baryon asymmetry is on or below the green band. 
We get the prediction that 
\begin{equation}
\laq{pred}
\L\lesssim 10~\text{--}~ 1000 \TEV
~{\rm with } ~ 
T_R = \O(1~ \text{--}~ 100) \GEV\,,
\end{equation}
for $B\xi_{CP}\lesssim \O(1).$
This is our main result.

We mention that at $t<t_R$ where the thermal plasma has even higher temperature $T>T_R$, 
the decay of $\f$ should also create the baryon asymmetry via the quark flavor oscillation. 
The contribution would be dominant for the baryon asymmetry in some parameter choices 
due to the high power of $T$ in $\G_{\rm BV}$. 
However the prediction of maximum asymmetry \Eq{maxas} does not change much 
even after including the extra contribution. 
This is because that the maximal baryon asymmetry generated at $T>T_R$ is diluted due to the entropy production to be
$\D_B^{\rm max}/s|_{T_R\to T} \times (T_R/T)^{5}$ at $t=t_R$, 
which is smaller than \Eq{maxas}.%
\footnote{Strictly speaking, one should replace the quark mass to be a thermal mass 
if $T\gtrsim 100\GEV$. If $E_{\rm cm}\gg \L$, one may need to calculate the asymmetry in a renormalizable UV model. 
In any case, the maximum amount of asymmetry produced at $t< t_R$ is diluted 
to be less than \Eq{maxas}. }

An interesting observation of the scenario is that the neutron-antineutron oscillation can be tested 
when $(\k_{1,2,3})^{-1/5}$ are of same order $\L$ for any choice of the flavor indices (right-panel of the Fig.\,\ref{fig:1}), {\it i.e.} with a general flavor structure. 
In fact, the rate of neutron oscillation can be within the experimental reach~\cite{Phillips:2014fgb,Milstead:2015toa, Frost:2016qzt, Hewes:2017xtr} 
if 
\begin{equation}
(\k_{1,2,3})_{111111}^{-1/5}\sim \L \lesssim 1000\TEV \,,
\end{equation}
where the subscript ``1" denotes the first generation.
The current bound is $\L\gtrsim \O(100)\TEV$~\cite{BaldoCeolin:1994jz, Abe:2011ky}. (See also Refs.\,\cite{Rao:1982gt, Buchoff:2012bm,Syritsyn:2016ijx} for theoretical calculations 
and uncertainties of the neutrino-antineutrino oscillation rate.) 
If one assumes a special UV model with such a specific flavor structure that $\k_{1,2,3}$ may be suppressed for the first generation,
the scenario becomes irreverent to the experiments. The baryogenesis is still possible. For example, one of the operators irrelevant to the experimental bounds can easily be $\L \ll\O(100\TEV)$ and generate enough baryon asymmetry. (See the left-panel of Fig.\,\ref{fig:1}.)

\section{Numerical simulation}

In this section we numerically confirm the mechanism by solving kinetic equations~\cite{Sigl:1992fn}.  
We focus on the density matrices of the left-handed up-type quarks for simplicity as follows 
\begin{equation}
\left( {\r}\right)_{ij} 
=
\int_{|{\bf p}|\sim m_\f/2}{d^3 {\bf p}\over (2 \pi)^3}
\, 
{\r_{ij}({\bf p},t) \over s }\,,
\ebq 
(\bar{\r})_{ij} 
=
\int_{|{\bf p}| \sim m_\f/2} {d^3 {\bf p}\over (2 \pi)^3}
\, 
{\bar{\r}_{ij}({\bf p},t) \over s }\,.
\end{equation}
Here we only consider the density matrices for the high energy (monochromatic) quarks with initial typical momentum of $m_\f/2$
produced by the $\f$ decays. 
The lower energy modes have a suppressed interaction rate for baryon number violation, and the effect is negligible.

The quantum evolution of the density matrices can be followed by solving the kinetic equations. (See Refs.\,\cite{Sigl:1992fn, Asaka:2011wq, Hamada:2018epb} for derivations of the equations. We use the convention in Ref.\,\cite{Hamada:2018epb}) 
The equations are 
\begin{align}
  i\frac{d \rho}{dt} 
  = 
  [\Omega , \rho] - 
  \frac{i}{2} \{ \Gamma^d, \rho \}\,,
\label{eq:kinK}
\end{align}
\begin{align}
  i\frac{d \bar{\rho}}{dt} 
  = 
  -[\Omega ,\bar{ \rho}] - 
  \frac{i}{2} \{ \Gamma^d, \bar{\rho} \}\,,
\label{eq:kinK2}
\end{align}
where 
\begin{equation} 
(\Omega)_{ij} =\delta_{ij} {m_i^2\over m_\f}\,.
\end{equation}
The destruction rates for quarks are given by
\begin{align}
\left( \G^d \right)_{ij} 
&=
C' \a_3^2  T \sqrt{\dfrac{2T}{m_\f}} \d_{ij}+ \left( \G_{\rm BV} \right)_{ij}\,.
\label{eq:GammaK}
\end{align}
The second term is 
\begin{equation}
\left( \G_{\rm BV} \right)_{ij} 
= 
\frac{C_{ij}}{4\pi\cdot (16\pi^2)^2} 
\frac{E_{\rm cm}^8}{\L^{10}}\times \frac{ 3\zeta{(3)}T_R^3}{2\pi^2}\,,
\end{equation} 
which corresponds to $\G_{\rm BV}$ discussed in the previous section, where $C_{ij}$ is a positive definite 3 by 3 hermitian matrix, whose each component is of order $10^4$ for a general flavor 
structure. 

\begin{figure}[t!]
\begin{center}  
\includegraphics[width=105mm]{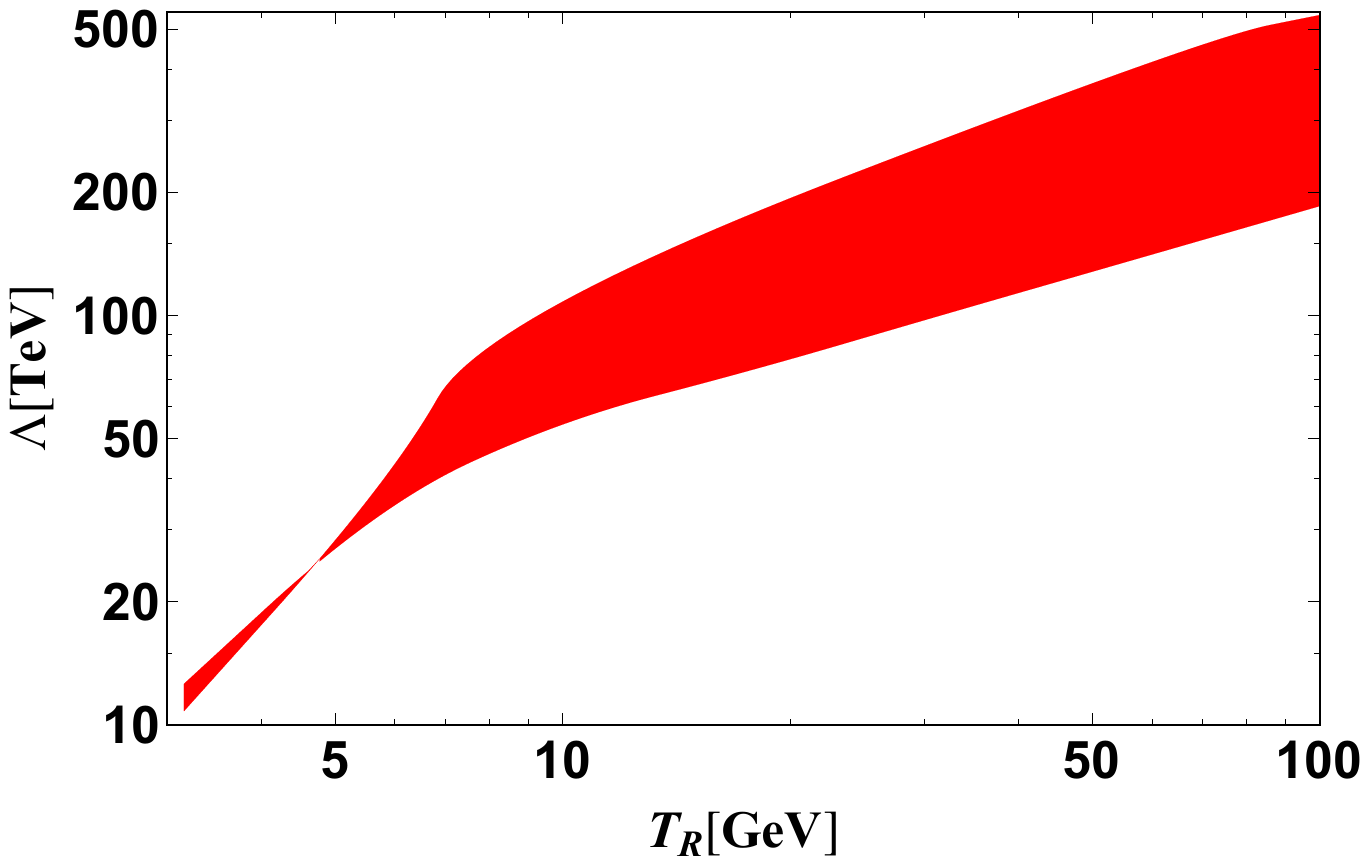} 
\end{center}
\caption{
The numerical result for the correct baryon asymmetry region in $T_R[{\rm GeV}]$-$\L [{\rm TeV}]$ plane. 
Here $2\L= E_{\rm cm}$, 
$V_i=1/\sqrt{3}\{\exp{(i)},\exp{(-2 i)},\exp{(i)}\}$ and $C_{ij}$ is randomly generated given in the main text, and $B=1$.
The range of the band corresponds to the variation of $C'=[0.3,3]$.   
}
\label{fig:2}
\end{figure}

\begin{figure}[t!]
\begin{center}  
\includegraphics[width=105mm]{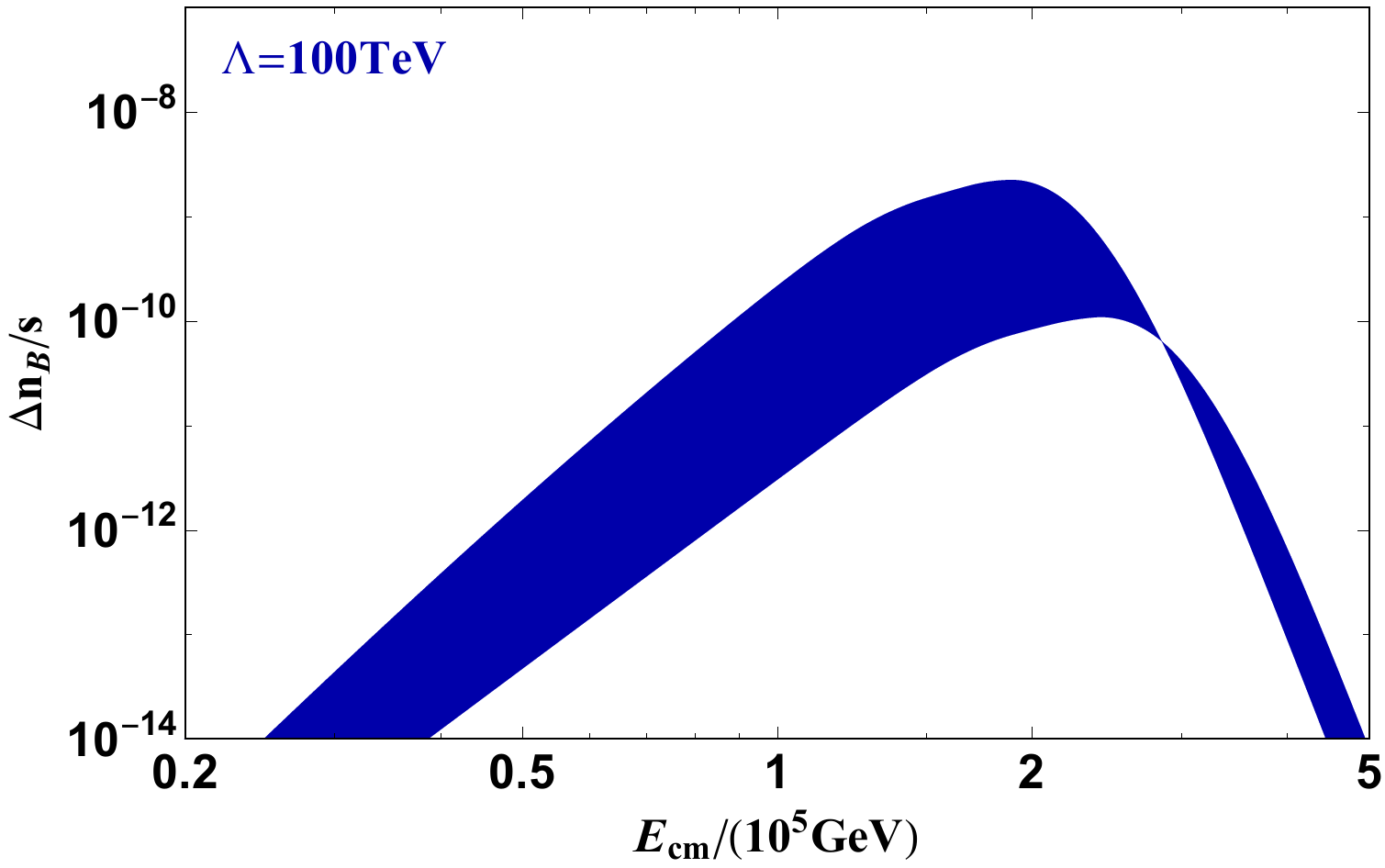} 
\end{center}
\caption{The numerical result for the baryon asymmetry generated by the first scatterings by varying $E_{\rm cm}=m_\f T_R$ with $T_R=90\GEV$ and $\L=100\TEV$. 
The other parameters are same as  Fig.~\ref{fig:2}. }
\label{fig:3}
\end{figure}

Now we are at the position to solve the kinetic equations. 
The initial condition can be given as 
\begin{equation}
\left( \r(t_R) \right)_{ij} 
= 
\left( \bar{\r}(t_R) \right)_{ij} 
= 
B {3\over 4}\frac{T_R}{m_\f}V_i^* V_j\,,
\end{equation}
which is provided by the direct decay of $\f$ at $t\simeq 1/\G_\f$. 
For simplicity, we have assumed that the decay product is a pure state.

Since the $2 \to 4$ interaction violates the baryon number by two units, 
we can estimate the generated baryon number by 
\begin{equation}
{\D n_B\over s} 
= 
2\int_{t=1/\G_\f}^{t=\infty}{dt\tr[(\r-\bar{\r}) \G_{\rm BV}] }\,.
\end{equation}
Notice again that we can neglect the baryon asymmetry production 
or destruction after the first scattering because $\G_{\rm BV}$ decreases significantly 
for lower center-of-mass energy.

In Fig.~\ref{fig:2} we show the numerical result of the allowed range 
in $T_R$-$\L$ plane with  $T_R m_\f=4\L^2$. 
Here we have also fixed $C_{ij}$ as
\begin{equation}
\left(
\begin{array}{ccc}
 5100 & 5000+3200 i & 4200+2900 i \\
 5000-3200 i & 10700 & 4400+200 i \\
 4200-2900 i & 4400-200 i & 5900 \\
\end{array}
\right)\,,
\end{equation}
which is generated at random, and  $V_i=1/\sqrt{3}\{\exp{(i)},\exp{( i)},\exp{(-2i)}\}$. The amount of the asymmetry as well as the behavior is consistent with the analytic estimation in the previous section.\footnote{One small difference is the behavior at low reheating temperature. 
When the $T_R$ is small, the produced baryon asymmetry gets suppressed since the oscillation becomes too fast. This effect is not discussed in the previous section. }
In Fig.~\ref{fig:3}, 
we show the produced baryon asymmetry at the first scatterings 
by varying $E_{\rm cm}$ with $T_R=90\GEV$ and $\L =100\TEV$. 
The other parameters are fixed to be the same as the previous figure. 
We find again that  there is an upper bound on the baryon asymmetry around $E_{\rm cm}\sim\bar{E}$. 
As noted, the baryon asymmetry evaluated with $E_{\rm cm } \gg \L \sim \O(0.1) \bar{E}$ may not be true due to the perturbativity bound. 
In such a high energy region, our effective theoretical treatment of the dimension nine operator becomes invalid. The baryon asymmetry of this region, however, can be re-evaluated in a UV renormalizable model.

\section{Case in the UV model}\label{sec:4}

One of the renormalizable UV models can be constructed by introducing two $Z_2$ even scalar quarks. 
(See Ref.~\cite{Aitken:2017wie} for another UV model.)
The interaction Lagrangian is given by
\begin{equation}
\laq{UVmodel}
{\cal L}^{\rm UV} \supset c_1\F_1 QQ+c_2 \F_2 d^* d^*+ c_3 \F_1 u^* d^*+ A \F_1^2 \F_2 +h.c.\,, 
\end{equation}
where $\F_1$ and $\F_2$ are heavy scalar quarks in the representations of $(-1/3,1,3)$ and $(2/3,1,3)$ under the SM gauge group.
For simplicity, we only consider one set of them, 
while the extension with multiple sets of scalar quarks is straightforward. 
$c_{1,2,3}$ ($A$) are the dimensionless (dimension 1) couplings.
By integrating out $\F_1$ and $\F_2$, which have the masses of $M_1$ and $M_2$ ($\gg \TEV$), respectively, 
one can obtain the dimension nine operators with the couplings, 
\begin{equation}
\laq{UVrele}
\k_1 \sim \frac{A^* c_1^2 c_2  }{ M_1^4M_2^2 }\,,~
\k_2 \sim \frac{A^* c_2 c_3^2  }{ M_1^4M_2^2 }\,,~
\k_3 \sim \frac{A^* c_1 c_2 c_3  }{ M_1^4M_2^2 }\,.
\end{equation}
For certain parameter choice, the model can be embedded into 
an $R$-parity violating supersymmetic model by identifying $\F_1 \AND \F_2$ with the superpartners of $d^*$ and $u^*$, respectively. (See {\it e.g.} Ref.~\cite{Barbier:2004ez}.)

Let us briefly discuss the baryogenesis in the context of the UV model. 
At low energy it has the same baryogenesis mechanism as in the previous section because of the decoupling theorem. 
In fact, in addition to the dimension nine operators, there are also dimension six operators,\,{\it i.e.} four-Fermi operators, by integrating out $\F_{1,2}.$
Since the flavor and CP violation are essential for our scenario, 
the four-Fermi operators may also have flavor- and CP- violating structures. 
The four-Fermi operators are generated as $\tl{G}^{d}_F (dd)^* dd$ by integrating out $ \F_2$ or as $\tl{G}^{[ud]}_F(ud)^* ud, \tl{G}^{[ud]}_F(QQ)^* ud, \OR\tl{G}^{[ud]}_F(QQ)^* QQ$ by integrating out $ \F_1.$
Here $\tl{G}^d_F \sim |c_2|^2/ M_2^2, \AND \tl{G}^{[ud]}_F \sim c_{1,3}^* c_{1,3}/M_1^2.$
The presence of the four-Fermi operators do not change our previous discussion on the baryogenesis significantly if $|\tl{G}^{d, [ud]}_F|\lesssim \O(|\k_{1,2,3}|^{2/5})$ since they do not generate/washout the baryon asymmetry  and do not contribute much on the thermalization.  

Some of the operators, however, could contribute to FCNC process in the ground-based experiment and $\tl{G}_F^{d}$ is constrained severely~\cite{Isidori:2010kg},
\begin{equation}
(\tl{G}^{d}_F)^{-1/2}>\O(10^2-10^5) \TEV\,,
\end{equation}
depending on the omitted indices for flavor and chirality. The type of $\tl{G}_F^{[ud]} (ud)^* ud$, on the other hand, is not severely constrained. 
These constraints can be avoided if $|c_{2}|$ are small enough and/or $\F_{2}$ are heavy enough.

Now let us estimate the maximum amount of the baryon asymmetry produced in the UV model due to the quark flavor oscillation.  
When we increase $E_{\rm cm}$ for given $T_R\AND \L$ in the effective theory by integrating out $\F_{1,2}$, 
$\G_{\rm BV}$ increases.
By assuming $M_1\sim M_2$ for simplicity, the increase of $\G_{\rm BV}$ continues until $E_{\rm cm}\simeq M_{1}+M_2$, 
where the exotic colored scalars, $\F_{1}\AND \F_2$, become on-shell. 
Thus the baryon number violating scattering is most efficient when $E_{\rm cm}\sim M_{1}+ M_{2}$
due to a $2\to 2$ process: {\it e.g.}
\begin{equation}
u+d \to \F_1^*+\F_2^*\,.
\end{equation}
The rate is given by 
\begin{equation}
\laq{BVcomp}
\G'_{ \rm BV} 
\sim 
\frac{3 |c_{1,3}|^2  |A|^2 }{ \, 4\pi E_{\rm cm}^4}  \times  \frac{3\zeta{(3)}T_R^3}{2\pi^2} 
~~( E_{\rm cm}\gtrsim M_{1}+M_2)\,.
\end{equation}
When $E_{\rm cm} \gg M_i$, 
the interaction rate of the baryon number violating process is suppressed.

The maximal asymmetry can be estimated similarly by replacing 
$\G_{\rm BV}$ with $\G_{\rm BV}'$ in Eqs.~\eq{th} and \eq{as}.
\begin{equation}
\({\D_B^{\rm max} \o s} \)^{\rm UV}\sim 3\times 10^{-10} B \xi_{CP}  C'^{-2} |c_{1,3}|^2 \left|\frac{A}{E_{\rm cm}}\right|^2 \(\frac{100\TEV}{E_{\rm cm}}\)^4  \({T_R\o 100\GEV}\)^{2}\,. 
\end{equation}
where  $E_{\rm cm} \sim \sqrt{T_R m_\f} \sim M_1+M_2$, 
where we again use the top quark mass for the dominant oscillation effect.
With $M_2\sim M_1\lesssim \O(100)\TEV$ the enough amount of baryon asymmetry 
can be generated with $A\sim M_1$, $c_{1} \OR c_3=\O(1)$. Interestingly it does not depend on $|c_2|.$ 
Thus, if the correct baryon asymmetry is produced by this process, 
the constraint from the neutron-antineutron oscillation as well as the FCNCs can be avoided by taking $|c_2|$ small enough.

Other than the testability in the neutron-antineutron oscillation experiments, 
this UV model may be tested in the flavor physics. 
Although the tree-level operators for the FCNC can be suppressed without conflicting with the baryogenesis, 
the operators are generated by box diagrams with Higgs/W-boson and $\F_1$ fields propagation.
Notice that $|c_{1}|$ or $|c_{3}|$, and $|1/M_1|$ should be large for the baryon asymmetry, and the loop contribution may be tested in 
future measurements of the FCNCs and CP-violating processes. $\F_{1,2}$ can be searched for in future hadron colliders.

\section{Discussion}

\paragraph{Estimation on $\phi$-coupling}
The interaction between $\f$ and quarks 
could be Planck-scale suppressed and is weak. This is the case if $\f$ is a modulus, string axion or a gravitino. 
The former two may also play the role of  the inflaton. 
The total decay width is $\G= g m_\f^3/4\pi M_{\rm pl}^2 $ with a model-dependent constant $g$. 
For instance, if $\f$ is a singlet scalar, ${\cal L}\supset\l \f H Q u/M_P$ 
represents the interaction to the SM model particles, 
where $\l$ is a dimensionless coupling constant.  
Then the dominant decay is to three-body final states with $g\sim \l^2/16\pi^2$ 
due to the phase space suppression (Our previous results do not change much by considering three-body decays instead of two-body.).
The reheating temperature is obtained as 
\begin{equation}
\laq{rh}
T_R 
\simeq 
100\GEV \cdot g^{1/2} \left( {m_\f\over 100\,{\rm PeV}} \right)^{3/2}\,.
\end{equation}
One can find that if the mass is smaller than $g^{-1/3} 100$\,PeV, 
the reheating temperature becomes smaller than the electroweak scale.  
For $1\GEV \lesssim T_R\lesssim 100\GEV,$ 
the center-of-mass energy of the emitted quarks at $t=t_R$
is around $2\TEV <g^{1/6} E_{\rm cm}\lesssim100\TEV$. 
Therefore in these kinds of models, the large enough center-of-mass energy 
can be realized consistently with our scenario. (See Figs.\,\ref{fig:1} and \ref{fig:2}.)

 Although we have considered $\f$ decays to quarks via higher dimensional operators, in the UV model $\f$ may also decay to $\F_i$ due to additional interaction terms such as 
$\d {\cal L}^{\rm UV} \supset -A_i \f |\F_i|^2- \l_{\f\F}  \f \F_1^2 \F_2$.  Then, there could also be baryon asymmetry produced through $\f \to \F_1\F_1\F_2, (\F_1\F_1\F_2)^*$ processes. The processes are both CP and baryon-number violating when the phases of $A$ and $\l_{\f \F}$ are misaligned. However, the contribution is at most  $|\d \D_B/s| \sim ({T_R\o m_\f })  \e $, similar to thermal leptogenesis~\cite{Fukugita:1986hr}. Here $\e\sim \({{\rm Loop~Factor}\o (4\pi)^3}\) \frac{\Im[\l_{\f \F} A^\*] A_i}{m_\f}  \times \G_\f^{-1}$, where 
$(4\pi)^{-3}$ comes from the 3-body phase space and ${\rm Loop~Factor}$ represents the loop effect, whose imaginary part is needed to obtain a ``strong phase" for a CP-violation. 
For instance, if $\G_\f \sim {A_i^2\o 4\pi m_\f} $, {\it i.e.} the dominant decays are $\f\to \F_i\F_i^*,$ $\e\lesssim \frac{\rm Loop~Factor}{(4\pi)^3} \frac{|\lambda_{\f\F} A|}{A_i}\ll\frac{\rm Loop~ Factor}{(4\pi)^2} \frac{|A|}{m_
\f},$ where we have used $\l_{\f\F}^2 m_\f/(64\pi^3)\ll A_i^2/(4\pi m_\f)$, namely the decays of $\f \to \F_1\F_1 \F_2$ are less frequent than the dominant one. 
In the parameter region of interest, $m_\f\sim \O(1-100)\,$PeV, $T_R\sim \O(1-100)\GEV$, $|A|\lesssim M_i\lesssim \O(100)\TEV,$ the contribution is suppressed.\footnote{This contribution might be important in the regime with smaller hierarchy between $T_R$ and $m_\f$, and $A$ and $m_\f$. }
On the other hand, the subsequent decays or scatterings of $\F_i$ from $\f$ produce quarks. These quarks undergo flavor oscillations and may generate baryon asymmetry. Such effect was taken account of with certain $\xi_{\rm CP}$ and $B$. 
Alternatively, with several flavors of the``squarks", flavor oscillations among them can also be important for baryogenesis.

\paragraph{\boldmath $ T_R\gtrsim 100\GEV$ or $T_R\lesssim 1\GEV$}

In the main part, we have focused on the range~\eq{range}. 
The extension to $T>100\GEV$ is straightforward 
by replacing the Higgs-induced quark masses by the thermal mass of the quarks in \Eq{as}. 
We have checked the success of the scenario.
In this case, $E_{\rm cm}$ can be increased due to higher temperature, 
and hence $m_\f$ can be smaller,%
\footnote{Even $T_R\gtrsim m_\f$ is possible enhancing the asymmetry production if the reheating is via the dissipation process, in which case the asymmetry production can be  enhanced~\cite{Eijima:2019hey}. The dissipation effect is important for reheating in the ALP inflation model~\cite{Daido:2017wwb, Daido:2017tbr, Takahashi:2019qmh}. 
} 
and $\L$ can be larger than those in the main part. The extension to $T_R< 1\GEV$ is also possible because the quark flavor oscillation is still important within the time scale of confinement $(\L_{\rm QCD} m_i/m_\phi)^{-1}$ after the heavy $\f$ decay. Here $\L_{\rm QCD}\simeq 0.2\GEV$ is the 
QCD scale. 
Within the time scale, one can ignore the effect of confinement and our scenario works 
as well with a proper estimation of $\G_{\rm th}$ by taking account of the quark scattering on hadrons.  
In this case, we may need  $\L < \O(10)\TEV$ and a special flavor structure to suppress the neutron-antineutron oscillation rate.

\section{Conclusions}
In this paper, we have shown a new mechanism for baryogenesis 
with the reheating temperature lower than the electroweak scale. 
In the effective theory approach, 
we have assumed the presence of the dimension nine baryon number violating operators 
which preserves the baryon parity. 
The high energy quarks produced from the decays of heavy particles, 
such as inflaton, moduli, or gravitino, undergo flavor oscillation, 
and lose the energy through the scattering with the ambient thermal plasma. 
Although the dimension nine operators are very weak at low energy satisfying the experimental constraints, 
during the energy loss processes of the quarks they are so efficient that 
the sufficient amount of baryon asymmetry is created. 
The scenario can be tied to the neutron-antineutron oscillation 
if all the dimension nine operators do not have a significant hierarchy in size, and thus can be confirmed. 
We also discussed in a UV model that the flavor/CP observables are searched for in the future. 
Our mechanism is compatible with various low-reheating temperature scenarios or low cutoff scale models.\footnote{There are interesting theoretical implications on scalar dark matter~\cite{Graham:2018jyp, Guth:2018hsa, Markkanen:2018gcw, 
Ho:2019ayl, Takahashi:2019pqf, Okada:2019yne, AlonsoAlvarez:2019cgw, Marsh:2019bjr} in the context of low-scale inflation which lasts long enough. Alternatively, the mechanism may be important in the context of the ``trans-Planckian censorship conjecture" which restricts the inflation energy density smaller than~$10^9\GEV$~\cite{Bedroya:2019snp, Bedroya:2019tba}. See also Refs.~\cite{Mizuno:2019bxy,Kadota:2019dol} for altering the bound. In both cases, an even lower reheating temperature is required. }

\section*{Acknowledgment}
H.I. and W.Y. thank Theoretical Particle Physics Group at Niigata University for kind hospitality when this work was initiated and partially done. 
The work of  T.A. was partially supported by JSPS KAKENHI Grant 
Numbers 17K05410, 18H03708, and 19H05097.
The work of H.I. was partially supported by JSPS KAKENHI Grant Numbers 18H03708. 
W.Y. was partially supported by NRF Strategic Research Program NRF-2017R1E1A1A01072736.

\providecommand{\href}[2]{#2}\begingroup\raggedright\endgroup

\end{document}